\documentclass[aps,prl,twocolumn,superscriptaddress,showpacs,
preprintnumbers,amsmath,amssymb]{revtex4}
\usepackage{graphicx}
\begin{document}

\title{Momentum dependence of orbital excitations in Mott-insulating titanates}
\author{C.~Ulrich}
\affiliation{Max-Planck-Institut~f\"{u}r~Festk\"{o}rperforschung,
D-70569 Stuttgart, Germany}
\author{L. J. P. Ament}
\affiliation{Institute-Lorentz for Theoretical Physics, Universiteit Leiden,
2300 RA Leiden, The Netherlands}
\author{G. Ghiringhelli}
\affiliation{CNR/INFM COHERENTIA and Dipartimento di Fisica, Politecnico di
Milano, 20133 Milano, Italy}
\author{L. Braicovich}
\affiliation{CNR/INFM SOFT and Dipartimento di Fisica, Politecnico di Milano,
20133 Milano, Italy}
\author{M. Moretti Sala}
\affiliation{CNR/INFM SOFT and Dipartimento di Fisica, Politecnico di Milano,
20133 Milano, Italy}
\author{N. Pezzotta}
\affiliation{CNR/INFM SOFT and Dipartimento di Fisica, Politecnico di Milano,
20133 Milano, Italy}
\author{T. Schmitt}
\affiliation{Swiss Light Source, Paul Scherrer Institut, CH-5232 Villigen PSI, Switzerland}
\author{G. Khaliullin}
\affiliation{Max-Planck-Institut~f\"{u}r~Festk\"{o}rperforschung,
D-70569 Stuttgart, Germany}
\author{J. van den Brink}
\affiliation{Institute-Lorentz for Theoretical Physics, Universiteit Leiden,
2300 RA Leiden, The Netherlands}
\affiliation{Stanford Institute for Materials and Energy Sciences,
SLAC, Menlo Park, CA 94025, USA}
\author{H.~Roth}
\affiliation{II. Physikalisches Institut, Universit\"{a}t zu K\"{o}ln,
50937 K\"{o}ln, Germany}
\author{T.~Lorenz}
\affiliation{II. Physikalisches Institut, Universit\"{a}t zu K\"{o}ln,
50937 K\"{o}ln, Germany}
\author{B.~Keimer}
\affiliation{Max-Planck-Institut~f\"{u}r~Festk\"{o}rperforschung,
D-70569 Stuttgart, Germany}

\date{\today}

\begin{abstract}
High-resolution resonant inelastic x-ray scattering has been used to determine
the momentum dependence of orbital excitations in Mott-insulating LaTiO$_3$
and YTiO$_3$ over a wide range of the Brillouin zone. The data are compared to
calculations in the framework of lattice-driven and superexchange-driven
orbital ordering models. A superexchange model in which the ex\-perimentally
observed modes are attributed to two-orbiton excitations yields the best
description of the data.
\end{abstract}


\maketitle

Valence electrons in transition metal oxides exhibit a large variety of
ordering phenomena that are associated with unusual macroscopic properties
\cite{Tok00}. In order to elucidate the origin of these properties,
experimental research beginning in the 1950's has systematically investigated
the relationship between the occupation of $d$-orbitals on metal ions and the
magnetic ordering pattern. The leading paradigm for the interpretation of
these data is the ``Goodenough-Kanamori" framework, which is based on the
assumptions that the orbital occupation is determined by electron-lattice
interactions alone, and that the relative orientation of orbitals on
neighboring sites controls the exchange interactions between magnetic ions
\cite{Tok00}. While experiments on most transition metal oxides are in
accordance with this paradigm, it has been called into question by recent data
on the Mott insulators LaTiO$_3$ and YTiO$_3$ with the seemingly simple
electron configuration 3$d^1$. Not only does the spin dynamics of these
compounds appear to defy a quantitative description in terms of the standard
lattice-driven orbital ordering scenario \cite{Kei00,Ulr02}, thermodynamic
data \cite{Che08} on LaTiO$_3$ have even been interpreted as evidence of a
novel spin-orbital liquid state \cite{Khal00} that is qualitatively incompatible
with this scheme. An alternative approach that emphasizes the many-body
superexchange coupling between spin and orbital degrees of freedom while
treating the local electron-lattice interaction as a perturbation, can explain
some of these experimental findings \cite{Khal00,Khal03}. Other experiments,
however, have been interpreted in terms of the rigid orbital ordering
pattern predicted by the Goodenough-Kanamori framework
\cite{Sol06,Sch05,Hav05,Moc03,Ito03,Pav04,Cwi03}. Further experimental work is
therefore required in order to discriminate between the competing theoretical
scenarios and to develop a comprehensive understanding of the electronic
structure of these prototypical orbitally degenerate Mott insulators.

Momentum-resolved spectroscopies, which associate specific energy scales
directly with the chemical bonding pattern, are among the most powerful
experimental probes of transition metal oxides. In particular, the momentum
$q$ dependence of low-energy ($\sim 10$ meV) spin excitations determined by
neutron spectroscopy \cite{Kei00,Ulr02} and high-energy ($\sim 1$ eV) charge excitations
determined by electron \cite{Fin01}, photoemission \cite{Dam03}, and x-ray \cite{xray} spectroscopies has yielded
deep insights into the electronic structure of Mott insulators. Recently,
orbital excitations with intermediate energies ($\sim 250$ meV) have been
detected by Raman \cite{Ulr06} and infrared \cite{Rue05} spectroscopies in
Mott-insulating titanates, but these methods are limited to $q=0$. Because of
the much larger photon wave vectors involved in the scattering process,
resonant inelastic x-ray scattering (RIXS) can in principle serve as a probe
of the momentum dependence of the orbital excitations
\cite{Ishi00,Fort08-2,Ulr08}. This is particularly informative for the
titanates, because of sharply divergent predictions by the two theoretical
scenarios. Whereas orbital excitations in the lattice-driven theory are weakly
dispersive because their energies are dominated by the local Jahn-Teller
coupling \cite{Hav05,Moc03,Pav04}, the superexchange model predicts
collective modes with a substantial dispersion \cite{Khal03}. Instrumental
constraints have, however, thus far precluded RIXS experiments on orbital
excitations with $q \neq 0$.

We have overcome these limitations at the newly developed ADRESS beamline at the
Swiss Light Source \cite{Ghi06,Beamline} and report RIXS data with photon energies
near the Ti $L_{2,3}$ absorption edge of LaTiO$_3$ and YTiO$_3$ over a large
fraction of the Brillouin zone. The experiments were performed on high-quality YTiO$_3$ and LaTiO$_3$ single
crystals (grown as described in Ref.~\onlinecite{Cwi03}) with magnetic
transition temperatures $T_C = 27$ K and $T_N = 146$ K, respectively. The
YTiO$_3$ (LaTiO$_3$) crystal was untwinned (partly twinned in the $ab$ plane
of the orthorhombic space group $Pbnm$). Since the orthorhombic distortion is
small, we use the pseudocubic notation, i.e. $a^\prime$ = $a / \sqrt{2}$ ,
$b^\prime$ = $b / \sqrt{2}$, and $c^\prime$ = $c/2$. The crystals were cut and
polished with surface normals along the [110] and [100] directions,
respectively, for LaTiO$_3$ and YTiO$_3$. This allowed investigations of
orbital excitations with momenta along these two directions in the Brillouin
zone. The RIXS data were taken at the ADRESS beamline, where an energy resolution of $55 \pm 2$~meV was
obtained using a combination of focusing elements and
high-precision optical gratings \cite{Ghi06,Beamline}. X--ray absorption spectra taken to select the energy for the RIXS
experiment were identical to those presented in Ref. \onlinecite{Ulr08}.
To study the momentum dependence of the orbital excitations, the incident
photon energy was adjusted to either 452.0~eV or 454.6~eV at the Ti L$_3$ edge
(corresponding to excitations from the $2p_{3/2}$ core level into the $3d$ $t_{2g}$
and $e_g$ levels, respectively), and the scattering angle $2\theta$ was
changed in a range between 50$^\circ$ and 130$^\circ$. Data were taken at
$T = 13$ K (i.e. in the magnetically ordered states) and at room temperature.
The incident x-ray beam was polarized either horizontally or vertically
(parallel or perpendicular to the scattering plane, respectively).
While excitations of comparable intensity were detected in both
polarization geometries, elastic scattering arising from static disorder,
surface effects, and partly from thermal diffuse scattering is strongly
enhanced for vertical polarization. The elastic peak in vertical polarization
was therefore used for calibrations of the photon energy (performed
periodically between inelastic scans) and a determination of the energy
resolution. The spectra shown below were taken in horizontal
polarization. They are sums of partial spectra, each accumulated for 10
minutes for a total of 4 hours.

\begin{figure}
\includegraphics[width=1.0\linewidth]{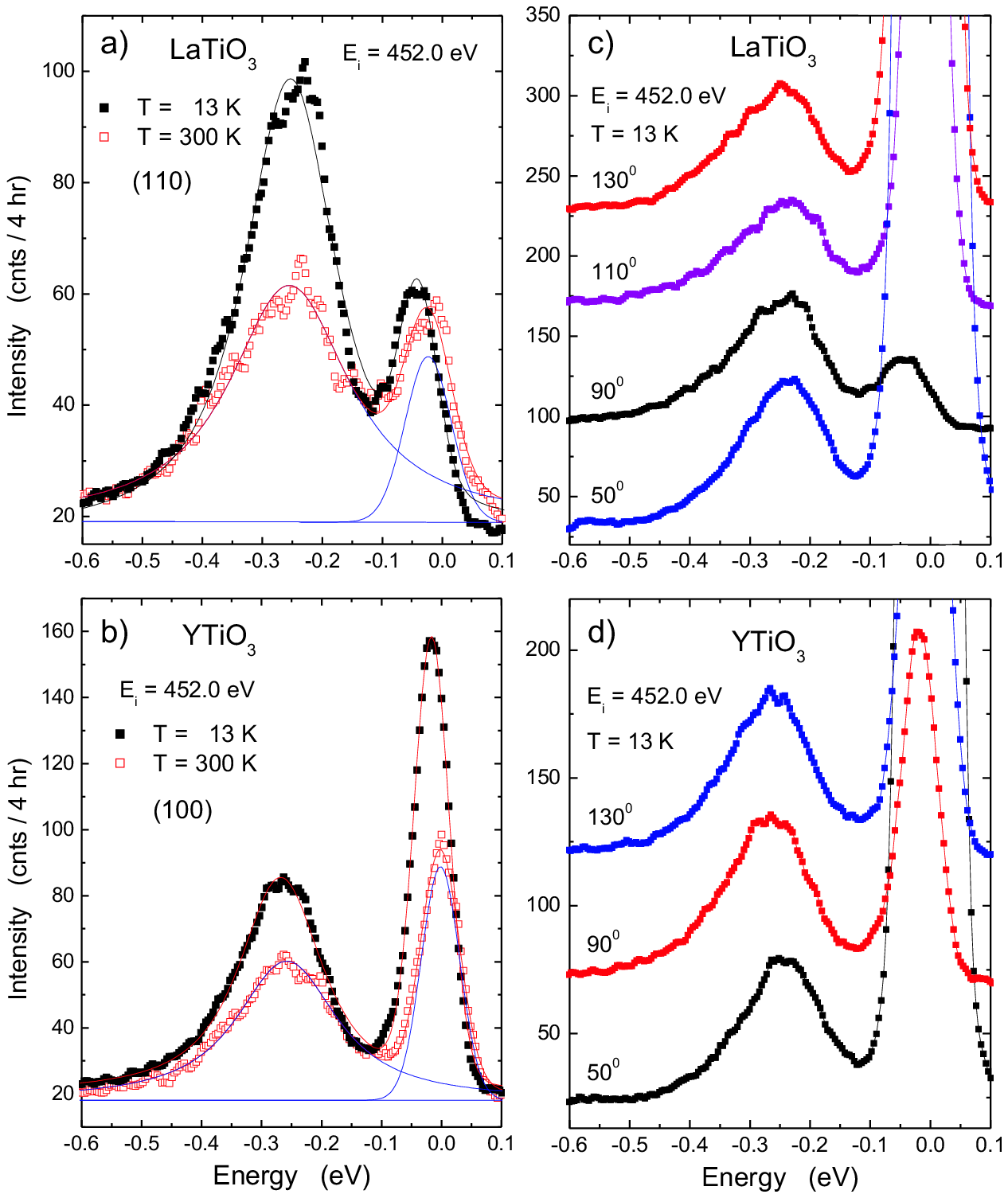}
\caption{\label{fig1} (Color online)
RIXS spectra of (a) LaTiO$_3$ and (b) YTiO$_3$ at room temperature and at
T = 13~K. The spectra were taken
at an incident energy E$_i$ = 452.0~eV (i.e. at the $t_{2g}$ level of the
L$_3$ excitation of the Ti--ions) in horizontal polarization of the incident
light at a scattering angle 2$\theta$ =~90$^\circ$. The lines are the results
of fits to the Voigt lineshape. (c,d) Evolution of the RIXS spectra as a
function of 2$\theta$. Note that the excitation momentum
$q = 2 k_i \sin \theta$, where $k_i$ is the incident photon wavevector.
The data along [100] ([110]) thus span a range of 11.5 to 25\% (8.5 to 17.5\%)
of the diameter of the Brillouin zone. The weak features
in the spectra arise from statistical fluctuations of the count rate,
which were smoothed out by the detector read-out algorithm.
}
\end{figure}

\begin{figure}
\includegraphics[width=1.0\linewidth]{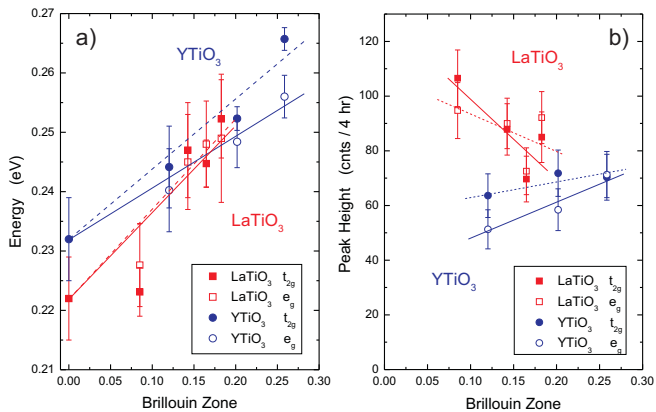}
\caption{\label{fig2} (Color online)
Momentum ($q$) dependence of (a) the energy and (b) the amplitude of
the RIXS signal of LaTiO$_3$ and YTiO$_3$ at T = 13 K. $q$ is given as a
fraction of the size of the Brillouin zone (BZ) along the high-symmetry
directions quoted in the text. Note that the BZ boundary is at 50\%.
The $q= 0$ data obtained from Raman light scattering \cite{Ulr06} are
also shown in (a). The lines are guides-to-the-eye.
}
\end{figure}

Figure 1 shows the $t_{2g}$ excitation spectra (incident photon energy 452.0
eV) of LaTiO$_3$ and YTiO$_3$ at room temperature and at $T = 13$ K in
horizontal polarization with $2 \theta = 90 ^\circ$ (panels a,b). In this
scattering geometry the diffuse elastic peak is strongly suppressed, and the
signal is dominated by inelastic scattering. All inelastic peaks appear on the
energy-loss side, as expected because the excitation energy of $\sim 250$ meV
greatly exceeds $k_B T$. The peaks disappear when the energy of the incident
beam is tuned away from the resonance condition, demonstrating that they arise
from excitations of the $3d$--electrons of the Ti ions.
The inelastic peaks sharpen with decreasing
temperature, and their intensities increase. The same effect has also been
observed by Raman light scattering \cite{Ulr06}. Interestingly, the ``elastic"
peaks at $T = 13$ K are not centered at zero energy, but at 40 (16) meV for
LaTiO$_3$ (YTiO$_3$). The shift closely corresponds to the energy range of
(anti-)ferromagnetic magnons previously detected by neutron
scattering \cite{Kei00,Ulr02}. At room temperature, where both compounds are
paramagnetic, the peak centers shift to 22 (0) meV, supporting an
interpretation in terms of magnons. With further improvement of the energy
resolution, RIXS may thus become widely applicable as a probe of dispersive
magnon excitations.

The momentum dependence of the RIXS spectra at $T = 13$ K is shown in
Fig.~1c,d. Kinematic and instrumental constraints restrict the $q$-range
to 18.2\% (25.8\%) of the Brillouin zone for YTiO$_3$ along [100] (LaTiO$_3$
along [110]). Note that the Brillouin zone boundary is at 50\%. In a first
step of a quantitative analysis, the data were fitted to Voigt profiles, which
correspond to a convolution of a Lorentzian with the Gaussian spectrometer
resolution (lines in Fig. 1a,b). The resulting peak
positions and amplitudes of both $t_{2g}$ and $e_g$ excitations are plotted in
Fig. 2, along with previously reported \cite{Ulr06,Ulr08} data for
$q=0$. Only a small variation of the peak positions is observed in the
$q$-range covered by our experiment (Fig. 2a). This behavior (which is  very
similar in both compounds) is incompatible with the strongly dispersive
single-orbiton excitations predicted by the superexchange model. Our
calculations below demonstrate, however, that the RIXS intensity of
two-orbiton excitations can exceed that of single-orbiton modes. The intensity
maximum of the two-orbiton continuum is determined by a convolution of two
single-orbiton states with momenta adding to $q$, and its dispersion is expected to average out.

In order to obtain a quantitative description of the contribution of
one- and two-orbiton excitations, we now consider the scattering
processes in detail. We first focus on a superexchange model for
YTiO$_3$ (Ref.~\onlinecite{Khal03}), and then compare the results to
those of a lattice-driven scenario. In RIXS, a $2p$ electron of the
Ti--ion is lifted to the $3d$ level,
modifying the electronic environment.
In particular, the core--hole potential effectively reduces the
Coulomb repulsion $U$ by the amount $U_c$, and therefore modulates
the superexchange interaction $J_{SE}=4t^2/U$. This exchange-bond
modulation mechanism was successfully used to explain the
observation \cite{Hill08} of two--magnon scattering by RIXS
\cite{Bri06,Ame07,Nag07}. In the titanates, modification of $J_{SE}$
causes either one or two orbital flips on neighboring Ti--sites.
Along the lines of Ref.~\onlinecite{Fort08-2}, the following
effective scattering operator can be derived:
\begin{eqnarray}
\hat{O}_{\bf q} = \sum\limits_{i,\delta} e^{i {\bf q} \cdot {\bf R}_i}
\left( J_1 \hat{n}^{(\gamma)}_{i+\delta}+J_2\hat{A}_{i,i+\delta}^{(\gamma)}
\right)\;,
\end{eqnarray}
where $J_1 = \frac{t^2}{U-U_c}-\frac{t^2}{U}$ and $J_2 = \frac{t^2}{U+U_c}
+ \frac{t^2}{U-U_c}-\frac{2t^2}{U}$. The operators
$\hat{n}^{(\gamma)}_i$ and $\hat{A}_{i,i+\delta}^{(\gamma)}$ were
taken from Ref.~\onlinecite{Khal03} and contain the density and
pseudospin one--half orbital operators whose explicit form depends
on the spatial direction $\gamma=a,b,c$ of the exchange bonds. The
first term $J_1$ represents single-site processes, whereas the
second term acts on two sites. In terms of orbitons, the latter term
only creates two orbitons, while the former can create both one and
two orbitons. The single-orbiton contribution scales with the
orbital order parameter which is strongly reduced \cite{Khal03} due
to the orbital frustration inherent in the superexchange model.
Therefore, the two-orbiton response
dominates the scattering intensity. Moreover, for the [100]--direction
the $J_1$-scattering channel is forbidden
for symmetry reasons, because the number of
$yz$-orbitals is conserved on each [100]--plane independently.

\begin{figure}
\includegraphics[width=1.0\linewidth]{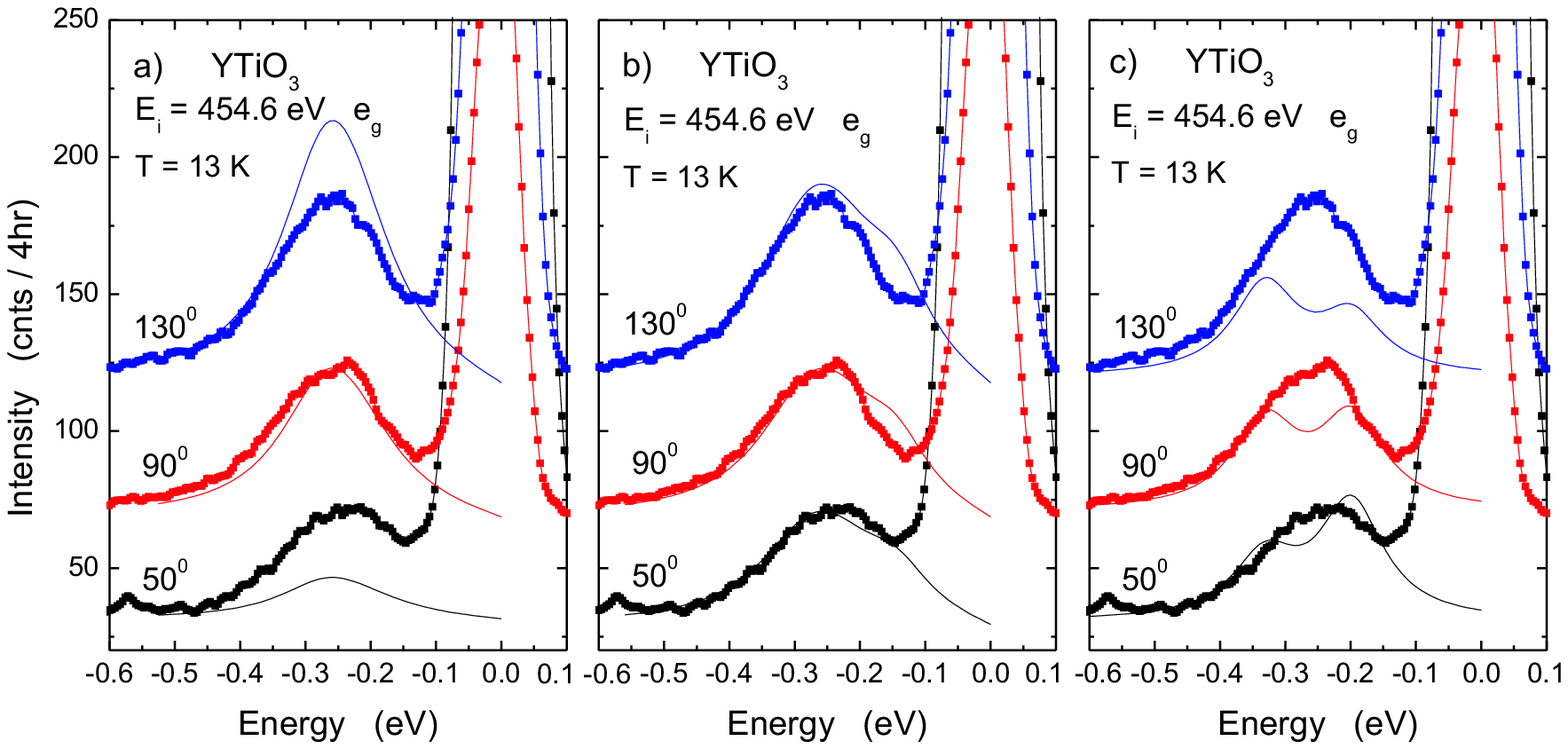}
\caption{\label{fig3} (Color online)
Comparison of the model calculations with the experimental RIXS data of
YTiO$_3$ for different scattering vectors $q$ along the [100]--direction:
(a) superexchange bond modulation,
(b) local shakeup of collective orbital excitations, and
(c) local crystal field excitation model.
}
\end{figure}

The results computed for the bond-modulation RIXS process are shown in
Fig. 3a. The scattering intensity was calculated using the orbital dispersion
relation from Ref.~\onlinecite{Khal03}, with the orbital exchange strength
$r_1J_{SE}=75$ meV (close to the value of 60 meV estimated
in Ref.~\onlinecite{Khal03}). In order to facilitate comparison with the
experimental data, we introduce an orbital damping $\gamma = 30$ meV due to
orbiton-orbiton (superexchange model) or orbital-phonon interactions
(Jahn-Teller model). The spectra are further convoluted with the resolution
function of the spectrometer. The lineshapes of the calculated and
measured RIXS spectra agree very well.

Figure 3a shows that the bond modulation mechanism alone fails to describe the
evolution of the absolute intensity of the RIXS signal with increasing $q$.
However, there is another coupling mechanism that also creates orbital
excitations. Namely, if the potential of the core--hole is not of $A_{1g}$
symmetry, it locally shakes up the occupied orbitals, which can leave the
Ti ions behind in an orbitally excited state. The corresponding
scattering operator can be written in a general form as
$\hat{O}_{\bf q}  = \sum_i \hat{O}_i \ e^{i {\bf q} \cdot {\bf R}_i}$.
Due to the octahedral symmetry of the TiO$_6$ unit, the local scattering
operator $\hat{O}_i$ can be decomposed into components
$(\hat{f}_{\kappa}^{(j)})_i$ transforming according to the rows
$\kappa$ of the irreducible representations $j$ of the octahedral group:
\begin{eqnarray}
\hat{O}_{\bf q} & = & \sum\limits_{j,\kappa} P_{j,\kappa} M_j
\sum\limits_i (\hat{f}_{\kappa}^{(j)})_i \ e^{i {\bf q} \cdot {\bf R}_i} \; .
\end{eqnarray}
The operators $(\hat{f}_{\kappa}^{(j)})_i$ induce an orbital transition of
certain symmetry, the polarization factors $P_{j,\kappa}$ depend on the
experimental geometry. The factors $M_j$ depend on the multiplet structure
of the intermediate states and can be obtained from cluster
calculations. We take only up to quadrupole components of
$(\hat{f}_{\kappa}^{(j)})_i$ into account (the higher multipoles contribute much less),
and assume that all $M_j$ are equal. The spectra
of the operators $(\hat{f}_{\kappa}^{(j)})_{\bf q}$ are calculated
within the superexchange model. Fig. 3b compares the result
of the calculations with the experimental data.
Here the energy scale $r_1J_{SE}$ was adjusted to 80 meV. Both
the lineshape and the scattering intensity of the observed inelastic peak are
reproduced for all scattering vectors. The only noticeable difference
is the weak shoulder that appears in the calculations at
about 120 meV. This arises from single orbiton excitations, which in this
case are not forbidden for the [100]--direction.

In addition to the orbital superexchange model, we calculated the
RIXS intensity in the opposite scenario
of strong orbital-lattice coupling. The orbital excitations are now local
crystal field transitions activated by the shakeup mechanism. We use
the orbital ordering pattern and crystal field splittings
(200 and 330 meV) of Ref. \onlinecite{Pav04} to obtain the spectra
shown in Fig. 3c. (Closely similar values were also obtained by other calculations \cite{Sch05,Rue05}.)
The crystal field transitions are localized and do
not show any dispersion, but due to the $q$-dependence of the transition
matrix element, the intensity ratio of the peaks can vary. With increasing
$q$, a shift of the overall spectral weight to higher
energies is thus seen, as in the experimental data. However, the calculated
spectrum is comprised of two well-separated peaks and is hence in poor
agreement with the RIXS data, which show a single broad feature. An ad-hoc
reduction of the level splitting away from the predicted values \cite{Sch05,Pav04,Rue05}
could improve the agreement of the lineshape, but this does not affect the $q$-dependence
of the intensity, which exhibits a trend opposite to what is observed.

In conclusion, we have determined the momentum dependence of the RIXS spectra
of LaTiO$_3$ and YTiO$_3$ and compared the results to quantitative predictions
of the superexchange and crystal-field models. We found that the former model
yields better agreement with the experimental data on YTiO$_3$. On a qualitative level, we note that the $q$-dependent
intensities of the orbiton excitations of YTiO$_3$ and LaTiO$_3$ (Fig. 2b) exhibit trends opposite to those of the
corresponding spin excitations \cite{Kei00,Ulr02}, as expected in superexchange models.
Mixing of spin and orbital
correlations by the superexchange interaction also accounts naturally for the
strong temperature dependence of the RIXS intensity, which is difficult to
explain in terms
of localized crystal field excitations.
An open problem is the spectral weight of single-orbiton excitations,
which is predicted to be larger than experimentally observed.
A possible explanation is that we have assumed the spins to be fully saturated,
while in YTiO$_3$ the ordered moment is lower than $1  \mu_B$ even at zero
temperature, presumably as a consequence of
joint spin-orbital fluctuations.
Such fluctuations are also expected to reduce the orbital order
parameter, to which the single-orbiton weight is proportional.

We thank M.W. Haverkort for useful discussions. L.A. thanks the MPI-FKF,
Stuttgart, for its hospitality. The crystal growth in Cologne was supported
by the DFG through SFB608. This work was performed at the ADRESS beamline of
the SLS (Paul Scherrer Institut) using the SAXES spectrometer developed jointly by
Politecnico di Milano, SLS, and EPFL.


\begin{thebibliography}{}
\bibitem{Tok00} For a review, see Y. Tokura and N. Nagaosa,
Science {\bf 288}, 462 (2000).
\bibitem{Kei00}  B. Keimer {\it et al.},
Phys. Rev. Lett.  {\bf 85}, 3946 (2000).
\bibitem{Ulr02} C. Ulrich {\it et al.},
Phys. Rev. Lett. {\bf 89}, 167202 (2002).
\bibitem{Che08} J.-G. Cheng, Y. Sui, J.-S. Zhou, J.B. Goodenough, and W.H. Su,
Phys. Rev. Lett. {\bf 101}, 087205 (2008).
\bibitem{Khal00}  G. Khaliullin and S. Maekawa,
Phys. Rev. Lett. {\bf 85}, 3950 (2000).
\bibitem{Khal03} G. Khaliullin and S. Okamoto,
Phys. Rev. Lett. {\bf 89}, 167201 (2002);
Phys. Rev. B {\bf 68}, 205109 (2003).
\bibitem{Cwi03} M. Cwik {\it et al.}, Phys. Rev. B {\bf 68}, 060401(R) (2003).
\bibitem{Sch05} R. Schmitz {\it et al.},
Phys. Rev. B {\bf 71}, 144412 (2005); Ann. Phys. (Leipzig) {\bf 14}, 626 (2005).
\bibitem{Hav05} M.W. Haverkort {\it et al.},
Phys. Rev. Lett. {\bf 94}, 056401 (2005).
\bibitem{Moc03} M. Mochizuki and M. Imada,
Phys. Rev. Lett. {\bf 91}, 167203 (2003).
\bibitem{Pav04} E. Pavarini {\it et al.},
Phys. Rev. Lett. {\bf 92}, 176403 (2004); New J. Phys. {\bf 7}, 188 (2005).
\bibitem{Sol06} I.V. Solovyev,
Phys. Rev. B {\bf 74}, 054412 (2006).
\bibitem{Ito03} T. Kiyama and M. Itoh, Phys. Rev. Lett. {\bf 91}, 167202
(2003).
\bibitem{Fin01} J. Fink {\it et al.}, J. Electron Spectrosc. Relat. Phenom. {\bf 117–-118}, 287 (2001).
\bibitem{Dam03} A. Damascelli {\it et al.}, Rev. Mod. Phys. {\bf 75}, 473 (2003).
\bibitem{xray} See, {\it e.g.}, Y.J. Kim {\it et al.}, Phys. Rev. B {\bf 76}, 155116 (2007).
\bibitem{Ulr06} C. Ulrich {\it et al.},
Phys. Rev. Lett. {\bf 97}, 157401 (2006).
\bibitem{Rue05} R. R\"{u}ckamp {\it et al.},
New J. Phys. {\bf 7}, 144 (2005).
\bibitem{Ishi00} S. Ishihara and S. Maekawa, Phys. Rev. B {\bf 62},
2338 (2000); S. Ishihara, Phys. Rev. B {\bf 69}, 075118 (2004).
\bibitem{Ulr08} C. Ulrich {\it et al.},
Phys. Rev. B {\bf 77}, 113102 (2008).
\bibitem{Fort08-2} F. Forte, L.J.P. Ament, and J. van den Brink,
Phys. Rev. Lett. {\bf 101}, 106406 (2008).
\bibitem{Ghi06} G. Ghiringhelli {\it et al.},
Rev. Sci. Instrum. {\bf 77}, 113108 (2006).
\bibitem{Beamline} For a description of the ADRESS beamline, see http://
sls.web.psi.ch/view.php/beamlines/adress/index.html
\bibitem{Hill08} J.P. Hill {\it et al.},
Phys. Rev. Lett. {\bf 100}, 097001 (2008);
L. Braicovich {\it et al.}, Phys. Rev. Lett. {\bf 102}, 167401 (2009).
\bibitem{Bri06} J. van den Brink and M. van Veenendaal,
Europhys. Lett. {\bf 73}, 121 (2006);
J. van den Brink, {\it ibid.} {\bf 80}, 47003 (2007).
\bibitem{Ame07} L.J.P. Ament, F. Forte, and J. van den Brink,
Phys. Rev. B {\bf 75}, 115118 (2007);
F. Forte, L.J.P. Ament, and J. van den Brink,
{\it ibid.} {\bf 77}, 134428 (2008).
\bibitem{Nag07} T. Nagao and J.I. Igarashi, Phys. Rev. B {\bf 75}, 214414 (2007).

\end{thebibliography}
\end{document}